\documentclass[a4paper,final]{eps}

\usepackage{graphicx}
\usepackage{times}

\volumenumber{xx}
\publishyear{200x}
\frompage{\pageref{firstpage}}
\topage{\pageref{finalpage}}

\shorttitle{M. Oka \lowercase{\textit{et al.}}: A GRADUAL ELECTRON EVENT}

\title{Non-thermal Electrons at the Earth's Bow Shock: A `Gradual' Event}
\author{M. Oka$^1$, T. Terasawa$^2$, M. Fujimoto$^{3}$, H. Matsui$^{4}$, Y. Kasaba$^5$, Y. Saito$^3$, H. Kojima$^6$, H. Matsumoto$^7$, and T. Mukai$^8$}
\affiliation{$^1$Kwasan Observatory, Kyoto University\\
             $^2$Tokyo Institute of Technology\\
             $^3$Institute of Space and Astronautical Science, Japan Aerospace Exploration Agency\\
	     $^4$University of New Hampshire\\
	     $^5$Tohoku University\\
	     $^6$Research Institute for Sustainable Humanosphere, Kyoto University\\
	     $^7$Kyoto University\\
	     $^8$Japan Aerospace Exploration Agency
	     }
\received{xxxx xx, 2008}\revised{xxxx xx, 2008}\accepted{xxxx xx, 2008}
\abstract{
Earth's bow shock is known to produce non-thermal electrons which are generally observed as a `spike' in their flux profile. Here, in this paper, we present an analysis of electron and whistler wave properties for a quasi-perpendicular shock crossing that is supercritical, but subcritical to the so-called whistler critical Mach number, M$^w_{\rm crit}$, above which  whistler waves cannot propagate upstream. We have found that the amplitudes of whistler waves increased exponentially as a function of time prior to the shock encounter, while the suprathermal ($>$ 2 keV) electron flux similarly increased with time, although with differing $e$-folding time scales. Comparison of the electron energy spectrum measured within the ramp with predictions from diffusive shock acceleration theory was poor, but the variation of pitch angle distribution showed scattering of non-thermal electrons in the upstream region. While not finding a specific mechanism to account for the electron diffusion, we suggest that the whistlers seen probably account for the differences observed between this `gradual' event and the `spike' events seen at shocks with no upstream whistlers.
}


\begin{document}
\label{firstpage}
\maketitle
\copyrighttext{}

\section{Introduction}
Energetic electrons with energies more than 20 keV have been observed at and near the Earth's bow shock (e.g. Fan et al., 1964; Frank and Van Allen, 1964; Anderson, 1965, 1969; Vandas, 1989). Since larger electron flux can be found on the interplanetary magnetic field (IMF) tangent to the bow shock (e.g. Anderson et al., 1979; Kasaba et al., 2000), electrons are considered to be accelerated in the quasi-perpendicular region where the shock angle $\theta_{Bn}$ is larger than 45$^{\rm o}$. Gosling et al. (1989) was the first to carry out comprehensive analysis of suprathermal ($<$20 keV) electrons across the shock front. In their quasi-perpendicular shock events, energetic electron flux was enhanced at the shock transition, and because of the localized feature, they termed their events as `spike' events. The energy spectrum showed a power-law form with the spectrum index of 3-4. The pitch angle distribution was almost isotropic at the transition layer while it was anisotropic in both the upstream and the downstream regions. More recently, Oka et al. (2006) conducted a statistical analysis of the power law indices measured in the shock layers. They reported that the power-law index of electron energy spectra is regulated by the so-called whistler critical Mach number M$^w_{\rm crit}$, which is defined as the critical point above which whistler waves cannot propagate upstream. 

In this paper, we report a shock crossing event that showed `gradual' profile of non-thermal electron flux in association with intensification of precursor whistlers. Contrary to the spike events reported by Gosling et al. (1989), the electron flux increased exponentially with decreasing distance from the shock. The event has been determined to be subcritical in relation to the whistler critical Mach number M$^w_{\rm crit}$. We will describe properties of the waves and discuss origin and transport of the non-thermal electrons in this gradual event.

% Their interpretation was that, because of very large thermal velocity, electrons are able to escape freely along background magnetic field line from a local acceleration site, leading to anisotropic distribution of T$_{\perp,e}>$T$_{||,e}$, where T$_{\perp,e}$ and T$_{||,e}$ are the perpendicular and the parallel temperatures of electrons, respectively. 

\section{Observation}
Our event is observed by Geotail at $\sim$03:10 UT on 11 February 1995 at an inbound crossing of the bow shock near the subsolar point, i.e. at (12.2, 4.0, 0.6) R$_{\rm E}$ in the GSE coordinate. Figure 1 shows the overview of main physical parameters of the gradual crossing event. The shock transition appears as an abrupt change in both the magnetic field data (MGF, Kokubun et al., 1994) as well as the plasma data (LEP, Mukai et al., 1994). The shock normal direction estimated by the Minimum Variance Analysis (Sonnerup and Cahill, 1967) was (0.94, 0.31, -0.17) consistent with the shock normal derived from the semi-empirical bow shock model of Peredo et al. (1995). This model is known to give normal directions in agreement with those obtained by the timing method of multi-spacecraft (Horbury et al., 2002). The upstream parameters, M$_A$ and $\theta_{Bn}$, were then estimated to be $\sim$6.8 and $\sim$68$^{\rm o}$, respectively. (For various methods of shock normal determination, see, e.g. Paschmann and Daly, 1998.)
\begin{figure}[t]
\centerline{\includegraphics[width=8.0cm,clip]{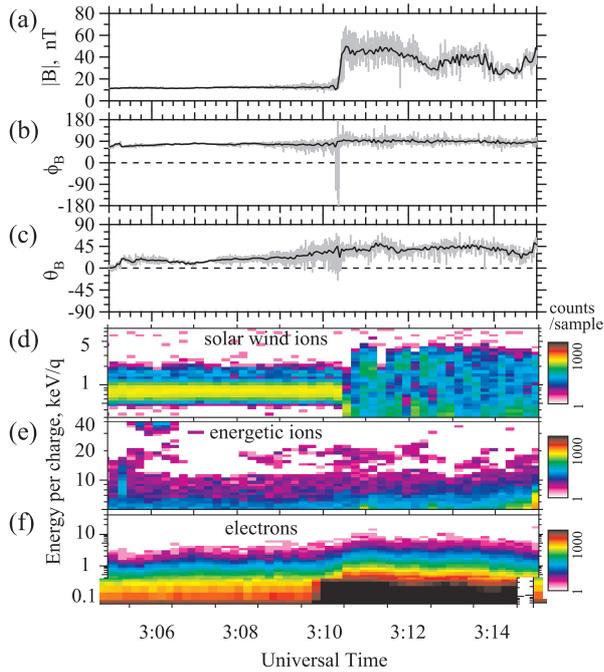}} 
\caption{From top to bottom are (a) magnitude $|$B$|$, (b) azimuthal $\phi_B$, and (c) latitudinal $\theta_B$ component of the magnetic field (MGF) with 3 sec (black line) and 1/16 sec (gray line) sampling, and the energy-time (Et) diagrams of (d) solar wind ions (LEP/SWI), (e) omni-directional energetic ions (LEP/EAI), and (f) omni-directional electrons (LEP/EAE), respectively.   }
\end{figure}
Noticeable in the magnetic field data in Figure 1 are the relatively large fluctuations increasing with time (03:08-03:10 UT, $\delta B/B \sim$20\%). Our particular interest also goes to the time profile of higher energy ($>$0.5 keV) electrons as can be seen in the panel (f) as a smooth increase of count rate in the time period of 03:09-03:11 UT. It is noted that the spacecraft soon exited from the magnetosheath at 03:15 UT. 

Figure 2 shows the detailed spectral properties of the waves. They are dominated by the right hand polarized component accompanied by a frequency cut off at around $f_{LH}$ which threshold seems not to be pointed out before.  We speculate that this is because whistler generation concerns both ion and electron dynamics. There are tens of detailed models for whistler generation and it is not the purpose of this letter to discuss the physical meaning of $f_{LH}$. From the Minimum Variance Analysis as well as the Means method (Means, 1972), the propagation angle $\theta_{kB}$ (the cone angle between $k$-vector and background magnetic field) were estimated to be 20-40{$^{\rm o}$}. We also removed the 180{$^{\rm o}$} ambiguity of the estimated $k$-vector using one component of electric field (EFD) data (Matsui et al., 1997). As a result, the waves with frequencies lower than $\sim$10 Hz indicated propagation toward the sun, away from the shock front, consistent with a past report (Orlowski et al., 1994). For higher frequencies, we could not obtain reliable results on the propagation direction, probably due to the low intensities of the waves. Note the cone angle between $k$-vector and the solar wind $V_{SW}$, $\theta_{kV}$, were estimated to be 60 - 90{$^{\rm o}$} so that the Doppler shift was not significant. Supportingly, the spectral slope of the high frequency range is approximately $\sim$ 5, consistent with past observations of right-hand polarized whistler waves (Orlowski et al., 1995). While traveling upwind, the waves suffered considerable (exponential) damping as shown in the left hand side of Figure 3 which shows the temporal profiles of band pass filtered magnetic field data. The characteristic time scale of the damping was calculated for each best fit model shown by the gray curves. This time scale was found to be 47 sec on average.

All observed features described above are well consistent with those of the so-called `1 Hz whistlers' reported elsewhere (e.g. Fairfield, 1974; Sentman et al., 1983;  Orlowski et al., 1994). From the above arguments, we conclude the observed waves to be the right hand polarized whistler waves propagating away from the shock front. 
%Note that 1 Hz whistlers are sometimes quoted as `upstream whistlers' to distinguish from the `foot whistlers' observed inside the shock transition (Orlowski et al., 1994). 

It is to be emphasized that, while thermal electrons had been studied with respect to whistler wave generation (e.g. Tokar et al., 1984; Orlowski et al., 1995), non-thermal electrons, to the author's knowledge, had never been observed in association with the upstream whistlers. Note again that the gradual profile of non-thermal electrons, has not been analyzed in past in connection with the bow shock crossing events as we will discuss below.

\begin{figure}[b]
\centerline{\includegraphics[width=8.0cm,clip]{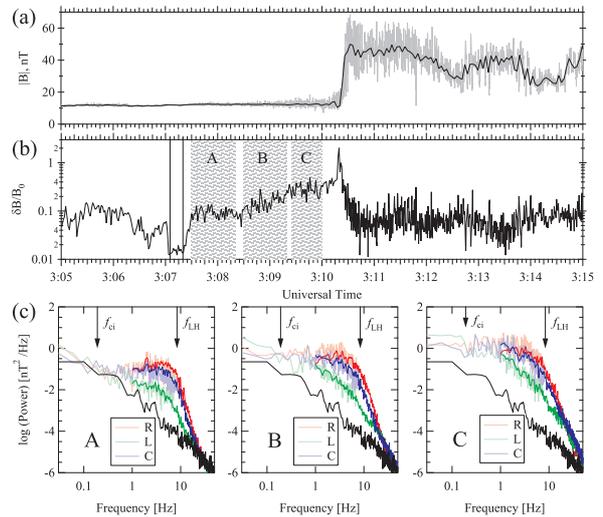}} 
\caption{From top to bottom are (a) the magnetic field magnitude (equivalent with Figure 1(a)), (b) root mean square of band-pass filtered magnetic field, (c) FFT spectra obtained at region A, B, and C indicated by the shaded regions in panel (b). The root mean square was obtained from frequency above 4 times the ion cyclotron frequency and has been normalized by the ambient magnetic field magnitude. Arrows in panel (c) show ion cyclotron frequency ($f_{ci}$=0.19Hz) and lower-hybrid frequency ($f_{LH}=\sqrt{f_{ci}f_{ce}}$=8.4Hz), whereas the black curve shows the background level obtained from 03:07:05 - 03:07:20 UT (indicated by two vertical lines). The red, green, and blue corresponds to the R, L, and C component from the fluxgate magnetometer (MGF/FX, $<$8Hz, light colored) and the search coil magnetometer (MGF/SC, $<$32Hz, heavy colored), respectively.}
\end{figure}

%Let us now describe the properties of the non-thermal electrons. 
The right hand side of Figure 3 shows the time profiles of electron phase space densities (PSDs). The flux increased exponentially as the spacecraft approached the shock front. The characteristic time scale of the increase was calculated for each best fit model shown by the red curves. Above 2 keV, the typical time scale was 24 sec.  During the flux increase, a pitch angle distribution also changed as shown in Figure 4. Black symbols indicate far upstream region of the shock and they show strong asymmetry (i.e. larger flux in $\mu<0$ region, where $\mu$ is the cosine of pitch angle $\alpha$), indicating that electrons were streaming away from the shock front. However, substantial amount of electron counts were detected in the $\mu>0$ region during the time interval from 03:08:30 to 03:10:00 UT as indicated by red symbols. By this time, the spacecraft was immersed in the precursor waves. The distributions isotropized in the immediate downstream (green and blue symbols). 

\begin{figure}
\centerline{\includegraphics[width=8.0cm,clip]{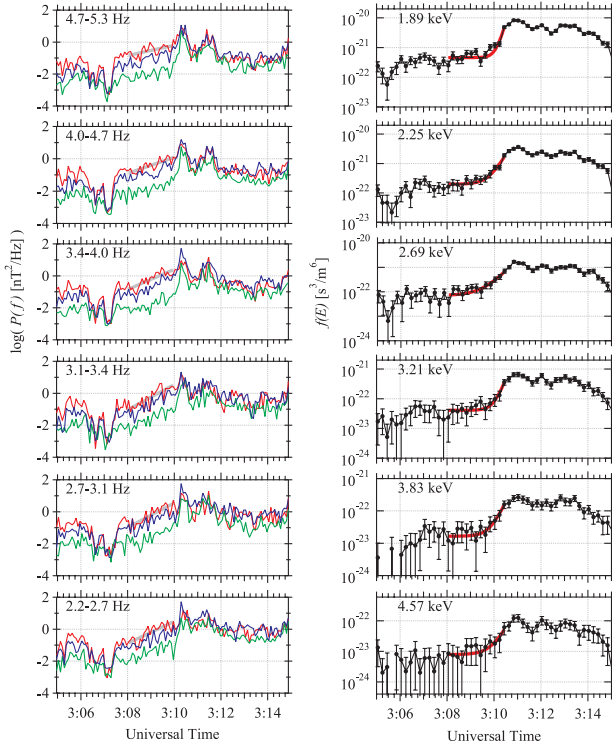}} 
\caption{(left) Temporal distributions of averaged power spectral densities (red, green, and blue for R, L, and C components) and their exponential fits (gray lines, 03:08:00 - 03:10:00 UT). Numbers in each panel show frequency ranges used for averaging. Frequencies are chosen so that they are the correspondence of the resonance conditions of electrons with energy shown to the right.  (right) Temporal distributions of electrons (black lines) and their exponential fits (red lines, 03:08:00 - 03:10:30 UT). Numbers in each panel show center energies of each energy channel.}
\end{figure}

Figure 5 shows the electron energy spectra. The gray line shows the spectrum obtained at 03:00 UT in the pure solar wind where there was no contamination from the bow shock. The open squares show the spectrum at 03:09:45 UT just prior to the crossing.  There was a significant amount of energetic ($>$1keV) electrons compared to the solar wind. The spectrum is roughly a power law. The filled squares show the spectrum at 03:10:26 UT just within the middle of the shock ramp. It now forms a complete power law above 2 keV. Then, we applied a chi-square fit above 2 keV with a power-law $f(E) \propto E^{-\Gamma} \exp{(-E/E_{\rm roll-off})}$, where $E$ is the electron energy as variable,  $\Gamma$=4.3($\pm$ 0.05), E$_{\rm roll-off}$=3.5($\pm$ 0.6), and the figures in the parenthesis are the 68\% confidence region. The resultant $\chi^2/d.o.f.$ was 28/21, where $d.o.f.$ is the degree of freedom and is equal to the number of data points minus the number of free parameters. A fit with the kappa distribution covering the whole energy range yielded relatively high $\chi^2/d.o.f.$.  Similar spectrum was observed in the immediate downstream as well. Note that E$_{\rm roll-off}$ was introduced to better fit the observation but we have not succeeded in deriving information of maximum attainable energy of electrons (Oka et al., 2006).

\begin{figure}
\centerline{\includegraphics[width=8.0cm,clip]{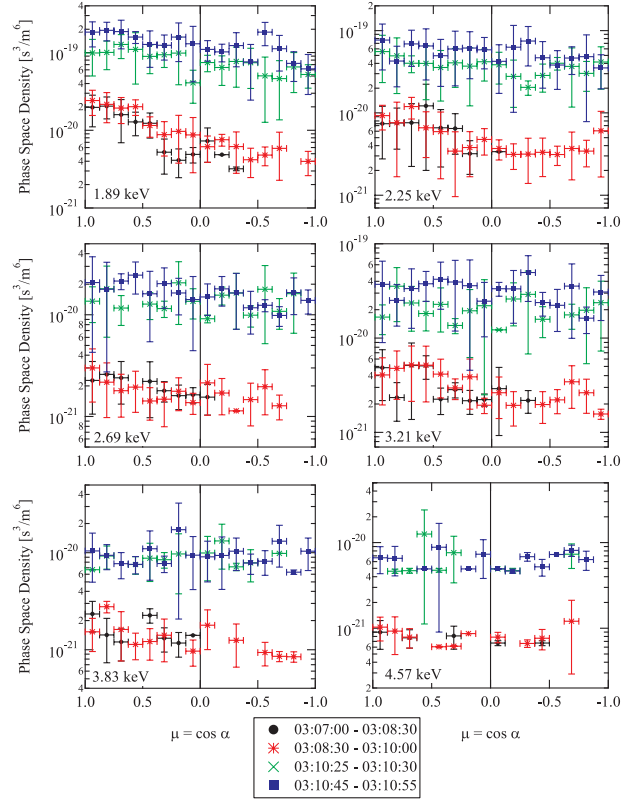}} 
\caption{Pitch angle distributions of electrons of energies, 1.89, 2.25, 2.69, 3.21, 3.83, and 4.57 keV. The vertical and horizontal axes show PSDs and $\mu$ (the cosine of pitch angle $\alpha$), respectively. Since the magnetic field was directed toward the sun, $\mu>0$ corresponds propagation away from the shock front. We have organized the obtained three dimensional distribution function by $\mu$ and the horizontal axis has been binned into 16 bins. The color code indicates different time intervals (in UT) as shown in the annotation.
}
\end{figure}

\begin{figure}
\centerline{\includegraphics[width=8.0cm,clip]{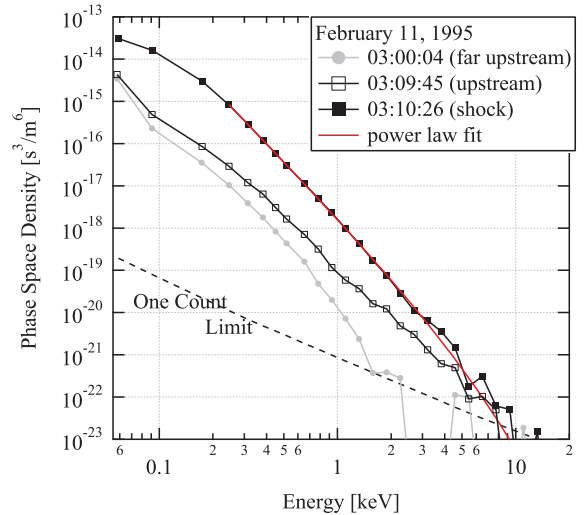}} 
\caption{Energy spectra of electrons during the bow shock transition.}
\end{figure}

\section{Discussion}

%The non-thermal electron flux in our event increased exponentially with decreasing distance from the shock front. 
%The `gradual' profile of electron flux as well as the indication of pitch-angle scattering are in complete contrast with the spike events reported by Gosling et al. (1989). In their event, isotropic, hard spectrum was obtained only in the vicinity of the shock ramp, and upstream electrons showed strong anisotropy indicative of scatter-free propagation away from the shock front.

The `gradual' profile of electron flux, the scattering of particles, and the appearance of power-law energy spectrum remind us of the diffusive shock acceleration (DSA, e.g. Blandford and Ostriker, 1978). In this process, particles are continuously scattered and move back and forth across the shock front to gain a net momentum from the velocity difference between upstream and downstream. However, there are some difficulties in the DSA to fully explain the observed features. The observed density ratio $r\sim2.5$ ($N_1=21$/cm$^{3}$ for the upstream and $N_2=52$/cm$^3$ for the downstream) and the classical DSA formula $\Gamma_{DSA}=3r/2(r-1)=2.5$ do not reproduce the observed $\Gamma\sim4.3$. The quantitative difference can readily be interpreted by, for instance, a free escape boundary that allows significant number of particles to escape from the acceleration region. The discrepancy between the $e$-folding distances derived from wave measurements (with the resolution of 47 sec) and particle measurements (with the resolution of 24 sec) indicates that the waves were not generated by the non-thermal electrons and that self-scattering was weak. Therefore, we do not consider the classic DSA mechanism to be fully responsible for generation of the observed power-law spectra.

% Nevertheless, we still consider there was substantial scattering in the upstream. The most widely used mechanism for scattering is the cyclotron resonance. Let us briefly estimate the scattering efficiency in terms of spatial diffusion coefficient $\kappa$ on the assumption of cyclotron resonance. It can be expressed as (Melrose, 1980)
% \begin{equation}
% \kappa = \frac{2v^2}{3\pi I(p) \Omega_{ce}} (G_+ + G_-)
% \end{equation}
% where $G_{\pm}=(1/\chi_o \mp 2\ln \chi_o - \chi_o)^{-1}$ are the geometrical factor of whistler wave propagation, $\chi=\cos(\theta_{kB})$ is the whistler cone angle, and $I(p)=f_RP(f_R)/B_0^2$ is the wave intensity calculated from the power spectral density $P(f)$ at the resonant frequency $f_R$. $\kappa$ strongly depends on $G_{\pm}$. By taking $\chi_o=\cos{50^{\rm o}}$, we can estimate the lower limit of $\kappa$ to be 10$^{18-19}$cm$^2$/s for the frequency range shown in Figure 3. This gives us the dimensionless intensity of scattering $\xi=\lambda_{||}/r_g$ to be 10$^{3-4}$ for electrons of a few keV where $\lambda_{||}$ is the mean free path and $r_g$ is the electron gyro-radius. Using an estimate of the shock normal speed in the observer's frame ($V_s\sim$250 km/s as derived roughly consistently from both particle flux and electric field conservation laws) and the $e$-folding time (24s), the characteristic length of the diffusion is calculated to be 6000 km which is smaller than $\lambda_{||}$(=1.5$\times$10$^{4-5}$km if $r_g$=15km) . Therefore, cyclotron resonance may not be efficient enough for the scattering.

Nevertheless, we still consider there was substantial scattering in the upstream as is evident from Figure 4. We have shown that the whistler waves were propagating along with the majority of the electrons. Since the direction must be opposite to that of electron streaming in order to satisfy cyclotron resonance condition, the cyclotron resonance by the right-hand-polarized whistlers is unlikely as the physical mechanism of scattering. The observed whistlers were oblique waves ($\theta_{kB}$=20-40$^{\rm o}$), and so the left hand polarized component and/or electrostatic component of the waves might have been playing a role in electron scattering. It is also to be mentioned that the amplitudes of the upstream waves were relatively high so that a non-linear effect should have played a role not only for the scattering but also acceleration. A recent theory indeed points out a possibility of efficient particle acceleration at a turbulent magnetic field where all scales are larger than the particle gyroradius (e.g. Giacalone, 2005; Jokipii and Giacalone, 2007). That the spectral index was relatively large also favors this theory, although the detailed comparison is left for future work.

We focus on the fact that the gradual profile was found at the shock with M$_{\rm A}$ slightly below M$^w_{\rm crit}$ (M$_{\rm A}$/M$^w_{\rm crit}\sim$0.65 in this event). Our interpretation is as follows. In a higher M$_{\rm A}$ shock (M$_{\rm A}>$M$^w_{\rm crit}$), whistler waves do not propagate upstream so that any electron that escapes the shock front cannot interact with the waves and that the intense flux can only be found at the shock front resulting in `spike' events. On the other hand, in a much lower M$_{\rm A}$ shock (M$_{\rm A}\ll$M$^w_{\rm crit}$), there would be plenty of waves to scatter particles, but the number of non-thermal electrons that are subject to scattering is small. Therefore, we expect to find similar events in the Mach number range slightly below M$^w_{\rm crit}$.

Unfortunately, however, the time resolution needs to be sufficiently high to resolve the decaying profile of non-thermal electrons. In the 78 events of Oka et al. (2006), there were a few events that seemed to be `gradual', but the decaying time was of the order of the time resolution of the particle measurement (12 sec), and thus we could not replicate the analysis presented in this paper. We anticipate that more sophisticated observations of multi-spacecraft mission such as Cluster, MMS, and SCOPE/CrossScale will reveal the nature of `gradual' events.

\acknowledgments
The authors are grateful to all members of the Geotail project. This work was partially supported by the Grant-in-Aid for Creative Scientific Research (17GS0208) from the MEXT, Japan. MO was supported by the Grant-in-Aid for JSPS Postdoctoral Fellows for Research Abroad.
%%\lastpagecontrol{20cm}

\email{M. Oka (e-mail: mitsuo.oka@uah.edu)}
\label{finalpage}
\lastpagesettings

\begin{references}\frenchspacing
%% Format for Journal Reference


% Anderson, K. A. et al., Energetic Electron Fluxes in and beyond the Earth's outer magnetosphere, \textit{J. Geophys. Res.}, \textbf{70}, 1039, 1965.

Anderson, K. A., Energetic Electrons of Terrestrial Origin Behind the Bow Shock and Upstream in the Solar Wind, \textit{J. Geophys. Res.}, \textbf{74}, 95, 1969.

Anderson, K. A., Intensity and Energy Spectrum of Electrons Accelerated in the Earth's Bow Shock, \textit{J. Geophys.}, \textbf{40}, 701, 1974.

Anderson, K. A. et al., Thin Sheets of Energetic Electrons Upstream From the Earth's Bow Shock, \textit{Geophys. Res. Lett.}, \textbf{6}, 401, 1979.

Blandford, R. D. and J. P. Ostriker, Particle acceleration by astrophysical shocks, \textit{Astrophys. J.}, \textbf{221}, L29-L32, 1978.

Fairfield, D. H., Whistler waves observed upstream from collisionless shocks, \textit{J. Geophys. Res.}, \textbf{79}, 1368-1378, 1974.

Fan, C. Y. et al., Evidence for $>$30keV electrons accelerated in the shock transition region beyond the Earth's magnetospheric boundary, \textit{Physical Review Letters}, \textbf{13}, 149, 1964.

Frank, L. A. and J. A. Van Allen, Measurements of energetic electrons in the vicinity of the sunward magnetospheric boundary with Explorer 14, \textit{J. Geophys. Res.}, \textbf{69}, 4923, 1964.

Giacalone, J., Particle acceleration at shocks moving through an irregular magnetic field, \textit{Astrophys. J.}, \textbf{624}, 765, 2005.

Gosling et al., Suprathermal Electrons at Earth's Bow Shock, \textit{J. Geophys. Res.}, \textbf{94}, 10,011-10,025, 1989.

Horbury et al., Four spacecraft measurements of the quasiperpendicular terrestrial bow shock: Orientation and motion, \textit{J. Geophys. Res.}, \textbf{107}, 273, 2002.

Jokipii, J. R. and J. Giacalone, Adiabatic compression acceleration of fast charged particles, \textit{Astrophys. J.}, \textbf{660}, 336, 2007.

Kasaba, Y. et al., Statistical studies of plasma waves and backstreaming electrons in the terrestrial electron foreshock observed by Geotail, \textit{J. Geophys. Res.}, \textbf{105}, 79-103, 2000.

Kokubun et al., The GEOTAIL Magnetic-Field Experiment, \textit{J. Geomag. Geoelectr}, \textbf{46}, 7-21, 1994.

Matsui, H. et al., Long-duration whistler waves in the magnetosheath: Wave characteristics and the possible source region, \textit{J. Geophys. Res.}, \textbf{102}, 17,583-17,593, 1997.

Means, J. D., Use of three-dimensional covariance matrix in analyzing the polarization properties of plane waves, \textit{J. Geophys. Res.}, \textbf{77}, 5551, 1972.

%Melrose, D. B., \textit{Plasma Astrophysics}, Gordon\&Breach, New York, 1980.

Mukai, T. et al., The Low Energy Particle (LEP) Experiment Onboard the GEOTAIL Satellite, \textit{J. Geomag. Geoelectr}, \textbf{46}, 669-692, 1994.

Oka, M. et al., Whistler critical Mach number and electron acceleration at the bow shock: Geotail observation, \textit{Geophys. Res. Lett.}, \textbf{33}, L24104, 2006.

% Orlowski, D. S. et al., Upstream waves at Mercury, Venus and Earth: Comparison of properties of one-Hertz waves, \textit{Geophys. Res. Lett.}, \textbf{17}, 2293-2296, 1990.

Orlowski, D. S. et al., On the source of upstream whistlers in the Venus foreshock, in \textit{COSPAR Colloquia, Plasma environments of non-magnetic Planets, vol. 4}, Edited by T. I. Gombosi, 217-227, Pergamon Press, New York, 1994.

Orlowski, D. S. et al., Damping and spectral formation of upstream whistlers, \textit{J. Geophys. Res.}, \textbf{100}, 17,117-17,128, 1995.

Paschmann, G. and P. W. Daly, Analysis Methods for Multi-Spacecraft Data, \textit{ISSI Scientific Report}, \textbf{SR-001}, pp.536, 1998.

Peredo, M. et al., Three-Dimensional Position and Shape of the Bow Shock and their Variation with {A}lfv\`{e}nic, Sonic and Magnetosonic Mach Numbers and Interplanetary Magnetic Field Orientation, \textit{J. Geophys. Res.}, \textbf{100}, 7907-7916, 1995.

Sentman, D. D. et al., The oblique whistler instability in the earths foreshock, \textit{J. Geophys. Res.}, \textbf{88}, 2048-2056, 1983.
 
Tokar, R. L. et al,. Whistler mode turbulence generated by electron beams in the bow shock, \textit{J. Geophys. Res.}, \textbf{89}, 105, 1984.

Vandas, M., Acceleration of Electrons by a Nearly Perpendicular Earth's Bow Shock: A Comparison Between Observation and Theory, \textit{Bull. Astron. Inst. Czech}, \textbf{40}, 175-188, 1989.









% (Author), (article's title), \textit{(Journal)}, \textbf{(Volume)}, (page number)--(page number), (year).
% 
% %% Format for Book
% (Author), (article's title), in \textit{(Book's title)}, Edited by (Editor), (total page) pp,
% (publisher), (published place), (year).
\end{references}
\end{document}